\documentclass [aps, prl, amssymb, twocolumn, showpacs]{revtex4-1}
\usepackage{graphicx,epsfig, amsfonts, amssymb, amsmath}
\usepackage{bm}
\usepackage{times}
\usepackage[colorlinks,citecolor=blue,linkcolor=red]{hyperref}
\usepackage{color}

\begin{document} 
 
\title{Predicted Rectification and Negative Differential Spin Seebeck Effect at Magnetic Interfaces}

\author {Jie Ren}\email{renjie@lanl.gov}
\affiliation{Theoretical Division, Los Alamos National Laboratory, Los Alamos, New Mexico 87545, USA} 

\date{\today}

\begin{abstract}
We study the nonequilibrium Seebeck spin transport across metal-magnetic insulator interfaces. The conjugate-converted thermal-spin transport is assisted by the exchange interaction at the interface, between conduction electrons in the metal lead and localized spins in the insulating magnet lead. We predict the rectification and negative differential spin Seebeck effect and resolve their microscopic mechanism, as a consequence of the strongly-fluctuated electronic density of states in the metal lead. The rectification of spin Peltier effect is also discussed. The phenomena predicted here are relevant for designing efficient spin/magnon diode and transistor, which could play crucial roles in controlling energy and information in functional devices.
\end{abstract}




\maketitle

Recently, spin Seebeck effect (SSE), a phenomenon that temperature bias can produce a spin current and an associated spin voltage, has been observed in magnetic metals~\cite{Uchida2008Nature}, semiconductors~\cite{Jaworski2010NatureMat, Breton2011Nature}, insulators~\cite{Uchida2010NatureMat, Uchida2010APL, Kikkawa2013PRL, Qu2013PRL} and even non-magnetic materials with spin-orbit coupling~\cite{Jaworski2012Nature}. Since then, the SSE has ignited a new upsurge of research interest, because it acts as a new method facilitating the functional use of heat and opens a new possibility of spintronics~\cite{spintronics}, magnonics~\cite{magnonics} and spin caloritronics~\cite{Slonczewski, BauerReview}. 

Of particular interest is the SSE in the insulating magnetic interface~\cite{Uchida2010NatureMat, Uchida2010APL, Kikkawa2013PRL, Qu2013PRL}. The reason is that different from spin-dependent Seebeck effect in metallic materials, SSE allows heat to generate a pure flow of spin angular momentum, a flow of spins {\it{without}} electron currents. This becomes obvious only after the observation of the SSE  through magnetic insulator-metal interfaces~\cite{Uchida2010NatureMat}. Itinerant electrons are often problematic in the thermal design of devices, of which the issue can be avoided by the SSE in magnetic insulator interfaces without conducting electron currents. It allows us to construct efficient thermoelectric devices upon new principles \cite{Kirihara2012NatureMat} and to realize non-dissipative information and energy transfer in the absence of Joule heating~\cite{Kajiwara2010Nature, TMI}.

In this Rapid Communication, we uncover interesting phenomena of the SSE across metal-insulating magnet interface: the {\it{rectification}} and {\it{negative differential}} SSE, that is, reversing the thermal bias gives asymmetric spin currents and increasing thermal bias abnormally gives a decreasing spin current. The rectification of spin Peltier effect (SPE) is also discussed.
We first demonstrate the absence of rectification and negative differential SSE in magnetic interfaces with {\it{constant}} electronic density of states (DOS) in the metallic lead. We then uncover that the {\it{strongly-fluctuated}} electron DOS is the key to retaining these intriguing spin Seebeck properties. As examples, we demonstrate the nontrivial rectification and negative differential SSE in several typical cases with non-smooth strongly-fluctuated electron DOS.
Our results readily render analytic interpretations and clear physical insights of the microscopic mechanism, which can provide further guidance for the optimization of the predicted effects in the future. 

The rectification and negative differential electronic/phononic conductances have played fundamental roles in realizing functional electronic/phononic diodes and transistors that are building blocks of modern electronics/phononics~\cite{phononics}. By the same token,  the predicted rectification and negative differential SSEs are also fundamental for constructing magnonic/spin caloritronic circuits with efficient spin Seebeck diodes and transistors. Therefore, we expect that our results would play crucial roles in spintronics~\cite{spintronics}, magnonics~\cite{magnonics} and spin caloritronics~\cite{BauerReview}, and could have potential applications in controlling energy and information in low-dimensional nanodevices~\cite{phononics}.

\begin{figure}
\hspace{-2mm}
\scalebox{0.35}[0.35]{\includegraphics{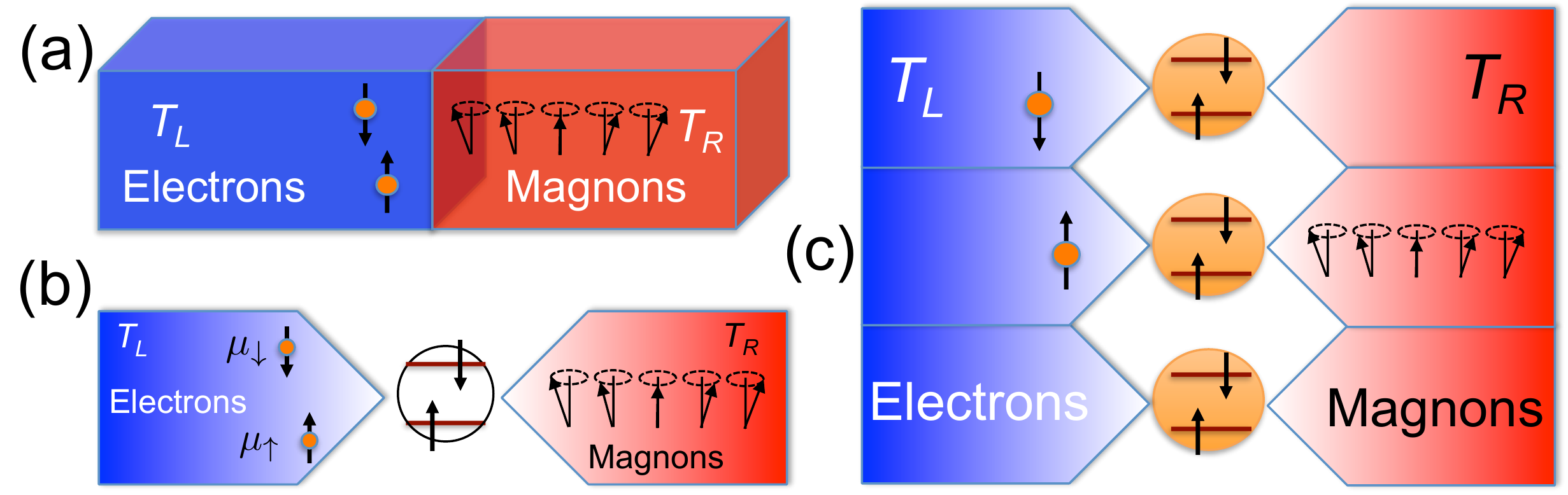}}
\vspace{-6mm}   
\caption{Schematic illustration of different setups of the metal-insulating magnetic interfaces.} 
\label{fig1}
\end{figure}

The magnetic interface system is schematically illustrated in Fig.~\ref{fig1}(a), similar to the setup of longitudinal spin Seebeck experiments~\cite{Uchida2010APL, Kikkawa2013PRL}. The left metallic lead is described by the free electrons: $H_L=\sum_{k\sigma}(\varepsilon_{k\sigma}-\mu_{\sigma})c^{\dag}_{k\sigma}c_{k\sigma}$, ($\sigma=\uparrow,\downarrow$), with possibly different spin-dependent chemical potentials $\mu_{\sigma}$ induced by spin accumulation \cite{Valet, Takahashi}. The spin voltage $\Delta\mu_s=\mu_{\downarrow}-\mu_{\uparrow}$ can be measured by the Hanle method \cite{Breton2011Nature} or be converted into an electric voltage through the inverse spin Hall effect \cite{Uchida2010NatureMat}. 
The right insulating magnetic lead can be described by a Heisenberg lattice $H_R=-J\sum_{\langle i,j\rangle}[{\frac{1}{2}S^+_iS^-_j+\frac{1}{2}S^-_iS^+_j}+S^z_iS^z_j]$, where $S^{\pm}_j$ is the raising (lowering) operator for the localized spin at site $j$, $S^{z}_j$ is the spin operator of the $z$ direction and  $J$ denotes the exchange coupling strength. The spin operators are conveniently mapped into bosonic magnons by Holstein-Primakoff transformation \cite{HP}: $S^{+}_j=\sqrt{2S_0-a^{\dag}_ja_j}a_j$, $S^{-}_j=a^{\dag}_j\sqrt{2S_0-a^{\dag}_ja_j}$, $S^z_j=S_0-a^{\dag}_ja_j$, where $S_0$ is the length of localized spins. At large spin limit or low temperatures ($\langle{a}^{\dag}_ja_j\rangle\ll2S_0$) we can approximate $S^{-}_j\approx\sqrt{2S_0}a^{\dag}_j$ and $S^{+}_j\approx\sqrt{2S_0}a_j$.  Clearly, the creation (annihilation) of a local magnon $a^{\dag}_j(a_j)$ at site $j$ corresponds to the lowering (rising) of the local spin component, i.e., excitation of magnons means that spins point less in the $z$-direction and magnetization goes down. Therefore, after a Fourier transform into the momentum space, the right insulating magnetic lead is approximated by the free magnon gas:
$H_R\approx\sum_{q}\hbar\omega_qa^{\dag}_qa_q+\text{const.}$,
where the dispersion of $\omega_q$ depends on the lattice details.

Similar to Refs.~\cite{Takahashi2010JPCS, Adachi2011PRB, Zhang2012PRB,Sothmann2012EPL}, the interfacial electron-magnon interaction is described by the $s$-$d$ exchange coupling~\cite{Mahanbook}:
\begin{equation}
H_{sd}=-\sum_{k,q} J_q[S^{-}_q c^{\dag}_{k\uparrow}c_{k+q\downarrow}+S^{+}_q c^{\dag}_{k+q\downarrow}c_{k\uparrow}],
\end{equation}
where $S^{-}_q\approx\sqrt{2S_0}\,a^{\dag}_q$, $S^{+}_q\approx\sqrt{2S_0}\,a_q$ are in the momentum space and $J_q$ denotes the effective exchange coupling at the interface. The first term $S^{-}_qc^{\dag}_{k\uparrow}c_{k+q\downarrow}$ describes the magnon emission process into the right insulating magnet associated with scattering of a spin-down electron to a spin-up electron at the left metallic side of the interface. The second term $S^+_qc^{\dag}_{k+q\downarrow}c_{k\uparrow}$ describes the reversed process that a spin-up electron near the interface absorbs a magnon from the right side and excites to a spin-down state with spin-flip. Note that the contribution of $-J_qS^{z}_q(c^{\dag}_{k\uparrow}c_{k\uparrow}-c^{\dag}_{k\downarrow}c_{k\downarrow})$ is customarily absorbed into $H_L$.

Considering each magnon carries a unit spin angular momentum of $-\hbar$ (associated with a magnetic moment), the magnonic spin current is equivalent to a spin-down current, which can be obtained by the Heisenberg equation $I_S=\frac{i}{\hbar}\langle[H_{sd},\sum_qa^{\dag}_qa_q]\rangle$.  Also, the magnon carries energy so that the magnonic heat current can be calculated through  $I_Q=\frac{i}{\hbar}\langle[H_{sd},\sum_q\hbar\omega_qa^{\dag}_qa_q]\rangle$. Following the approach as detailed in the Supplemental Material~\cite{Supple}, we obtain the spin and heat currents from left to right, respectively:
\begin{eqnarray}
I_S&=&\frac{2S_0}{\hbar}\int^{\infty}_0 d\omega F_R(\omega) \int^{\infty}_{-\infty} d\varepsilon \rho_L(\varepsilon) \mathcal W(\varepsilon, \omega),      \label{eq:bulkIS}\\
I_Q&=&\frac{2S_0}{\hbar}\int^{\infty}_0 d\omega F_R(\omega) \hbar\omega \int^{\infty}_{-\infty} d\varepsilon \rho_L(\varepsilon) \mathcal W(\varepsilon, \omega),    
\label{eq:bulkIQ}
\end{eqnarray}
with 
\begin{eqnarray}
\mathcal{W}(\varepsilon,\omega)&=&f_{L\downarrow}(\varepsilon+\hbar\omega)[1-f_{L\uparrow}(\varepsilon)][1+N_R(\hbar\omega)]   \nonumber\\
&-&f_{L\uparrow}(\varepsilon)[1-f_{L\downarrow}(\varepsilon+\hbar\omega)]N_R(\hbar\omega),
\label{eq:W}
\end{eqnarray}
where $\rho_L(\varepsilon)$ denotes the electron DOS in the left metal side; $F_R(\omega)$ contains the magnon DOS and the electron-magnon coupling strength, which is reminiscent of Eliashberg function~\cite{Mahanbook} in the field of electron-phonon scattering;
$f_{L\sigma}(\varepsilon)=[e^{{(\varepsilon-\mu_{\sigma})}/{(k_BT_L)}}+1]^{-1}$ is the electron distribution with spin $\sigma$ in the left metal side that is equilibrium at temperature $T_L$ and $N_R(\hbar\omega)=[e^{{\hbar\omega}/{(k_BT_R)}}-1]^{-1}$ denotes the magnon distribution in the right magnetic insulator that is equilibrium at temperature $T_R$.  Equation~(\ref{eq:bulkIQ}) is reminiscent of the thermal transport induced by interfacial electron-phonon coupling, studied in Ref.~[\onlinecite{Ren2013PRB}].

It is clear that the first product $f_{L\downarrow}(1-f_{L\uparrow})(1+N_R)$ in $\mathcal{W}(\varepsilon,\omega)$ describes the down-scattering rate of the occupied spin-down state with high energy $\varepsilon+\hbar\omega$ flipping to the empty spin-up state with low energy $\varepsilon$, accompanied by emitting a magnon with energy $\hbar\omega$ into the right magnet. The second product $f_{L\uparrow}(1-f_{L\downarrow})N_R$ reversely describes the up-scattering rate of the occupied low energy spin-up state flipping to the empty high energy spin-down state, with absorbing a magnon from the right magnet. These two spin-flip scattering processes, accompanied by the magnon emission/absorption, conserve not only the energy but also the spin angular momentum.

Following the procedure in the field of electron-phonon scattering, it is customary and legitimate to take a constant bulk electron DOS $\rho_L(\varepsilon){\approx}C_{L}$, because for a good metal  the electron DOS is flat and the integral over $d\varepsilon$ converges within a thermal energy $k_BT_L$ around the chemical potential~\cite{Mahanbook}. Therefore, by applying the equality
${\int}d\varepsilon f_{L\downarrow}(\varepsilon+\hbar\omega)[1-f_{L\uparrow}(\varepsilon)]=(\hbar\omega-\Delta\mu_s)N_L(\hbar\omega-\Delta \mu_s)$, 
Eqs.~(\ref{eq:bulkIS}, \ref{eq:bulkIQ}) finally lead to the formulas
\begin{eqnarray}
I_S&=&\frac{2S_0C_L}{\hbar}  \int^{\infty}_0 d\omega F_R(\omega) (\hbar\omega-\Delta \mu_s)  \nonumber \\
&\times& \left[N_L(\hbar\omega-\Delta \mu_s)-N_R(\hbar\omega)\right];   
\label{eq:bulkIS1} \\
I_Q&=&\frac{2S_0C_L}{\hbar}  \int^{\infty}_0 d\omega F_R(\omega) \hbar\omega (\hbar\omega-\Delta \mu_s)  \nonumber \\
&\times& \left[N_L(\hbar\omega-\Delta \mu_s)-N_R(\hbar\omega)\right].  
 \label{eq:bulkIQ1}
\end{eqnarray}
The expressions are familiar from earlier discussions with interfacial fermion-boson coupling~\cite{Mahanbook,Ren2013PRB,Mahan2009PRB,Bender2012PRL}.
We first examine the SPE that generates heat current from merely spin voltage, by taking $T_L= T_R$ but $\Delta\mu_s\neq0$.
Clearly, in the absence of thermal bias, the heat current $I_Q$ is asymmetric under reversing the spin voltage $\Delta\mu_s\rightarrow-\Delta \mu_s$ so that {\it{the rectification of SPE exists}}. Similarly, the spin current $I_S$ is also asymmetric under revering the spin voltage. 
Moreover, it is straightforward to prove that $\partial_{\Delta \mu_s}I_Q$ and $\partial_{\Delta \mu_s}I_S$ are both positive, thus {\it{the negative differential SPE and spin conductance are absent}}. 
By taking $\Delta \mu_s=0$ but $T_L\neq T_R$, we then examine the SSE that generates spin current from merely thermal bias, which is of our central interest. In this way, Eqs. (\ref{eq:bulkIS1}, \ref{eq:bulkIQ1}) reduce to Landauer-type formulas 
\begin{eqnarray}
I_S  \!&=&\! \frac{2S_0C_L}{\hbar}  \!\!\!  \int^{\infty}_0 \!\!\!\!\! d\omega F_{\!R}(\omega) \hbar\omega \left[N_{ \!L}(\omega) \!-\! N_{ \!R}(\omega)\right];    \label{eq:bulkIS2}\\
I_Q  \!&=&\! \frac{2S_0C_L}{\hbar}  \!\!\! \int^{\infty}_0 \!\!\!\!\! d\omega F_{ \!R}(\omega) (\hbar\omega)^2  \! \left[N_{ \!L}(\omega) \!-\! N_{ \!R}(\omega)\right]\!. \label{eq:bulkIQ2}
\end{eqnarray}
Since the temperature dependence only manifests in the Bose-Einstein distributions $N_L$ and  $N_R$, one can readily
prove that {\it{in this magnetic interface with constant electron DOS we can never have the rectification and negative differential SSE}}.

As we can see from above derivations, the constant electron DOS is the key assumption that leads to the Landauer-type Eqs.~(\ref{eq:bulkIS2},\ref{eq:bulkIQ2}). Therefore, {\it{to obtain the nontrivial rectification and negative differential SSE, we need metallic materials with strongly-fluctuated electron DOS}}~\cite{phonon_function}.

\begin{figure}
\scalebox{0.30}[0.30]{\includegraphics{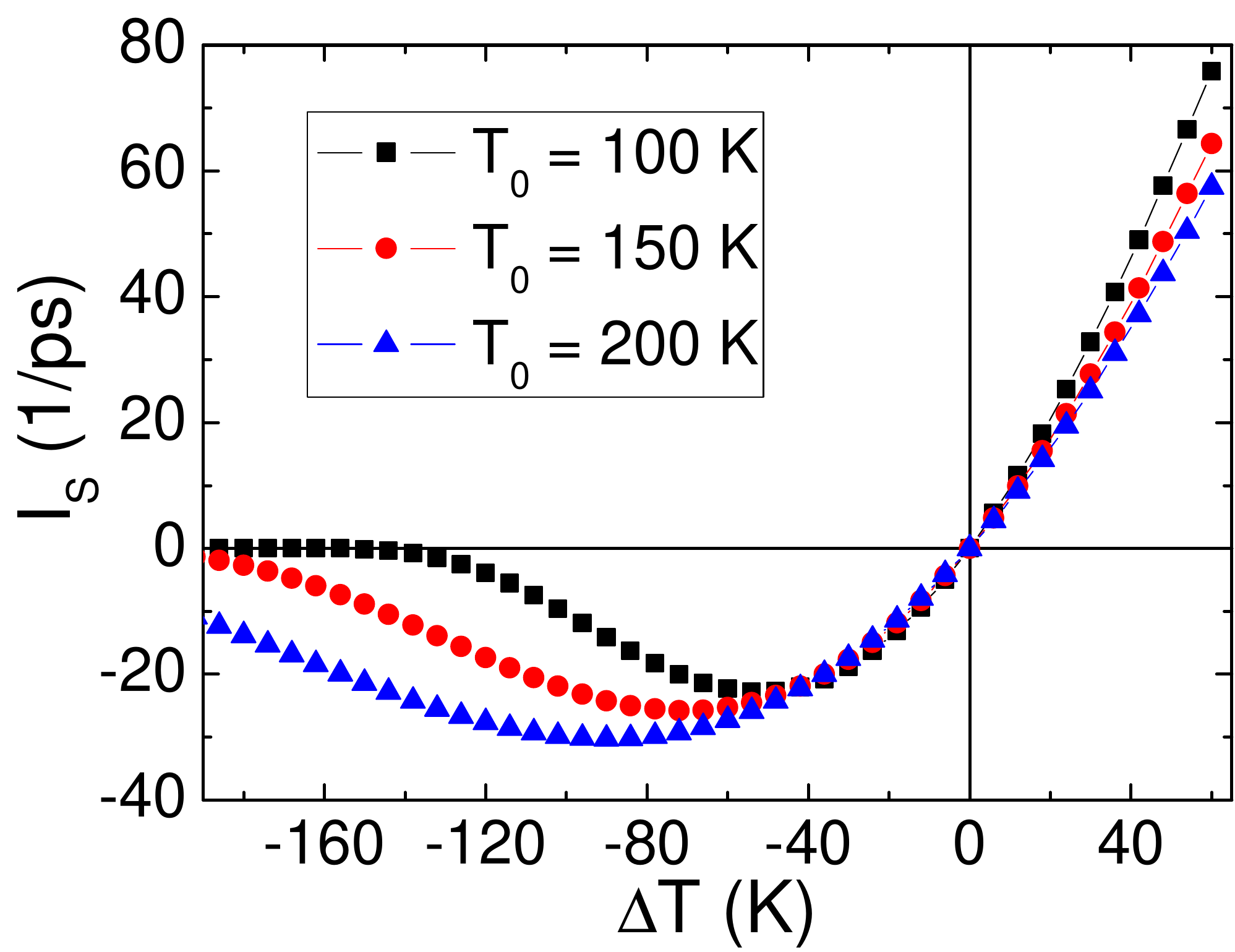}}
\vspace{-4mm}   
\caption{{\bf~Rectification and negative differential spin Seebeck effect in macroscopic magnetic interface}. Spin Seebeck current $I_S$ as a function of the temperature bias $\Delta{T}$ with $T_{L,R}=T_0\pm\Delta{T/2}$, for varying $T_0$ and Lorentzian peak positions: $T_0=100$~K, $\varepsilon_0=20$~meV; $T_0=150$~K, $\varepsilon_0=35$~meV; $T_0=200$~K, $\varepsilon_0=50$~meV. We set $S_0=16$, which is comparable with the experimental ones in typical magnetic insulators, cf. Refs.~\onlinecite{Adachi2011PRB,para1,para2}. $\mu_0=0$ and the Ohmic spectrum $F_R(\omega)=\alpha\frac{\omega}{\omega_c}e^{-\omega/\omega_c}$ is adopted, with $\alpha=10$, $\omega_c=50$~meV. Other choices of $F_R(\omega)$ will not change the results qualitatively. 
} \label{fig2}
\end{figure}

\begin{figure}
\scalebox{0.43}[0.43]{\includegraphics{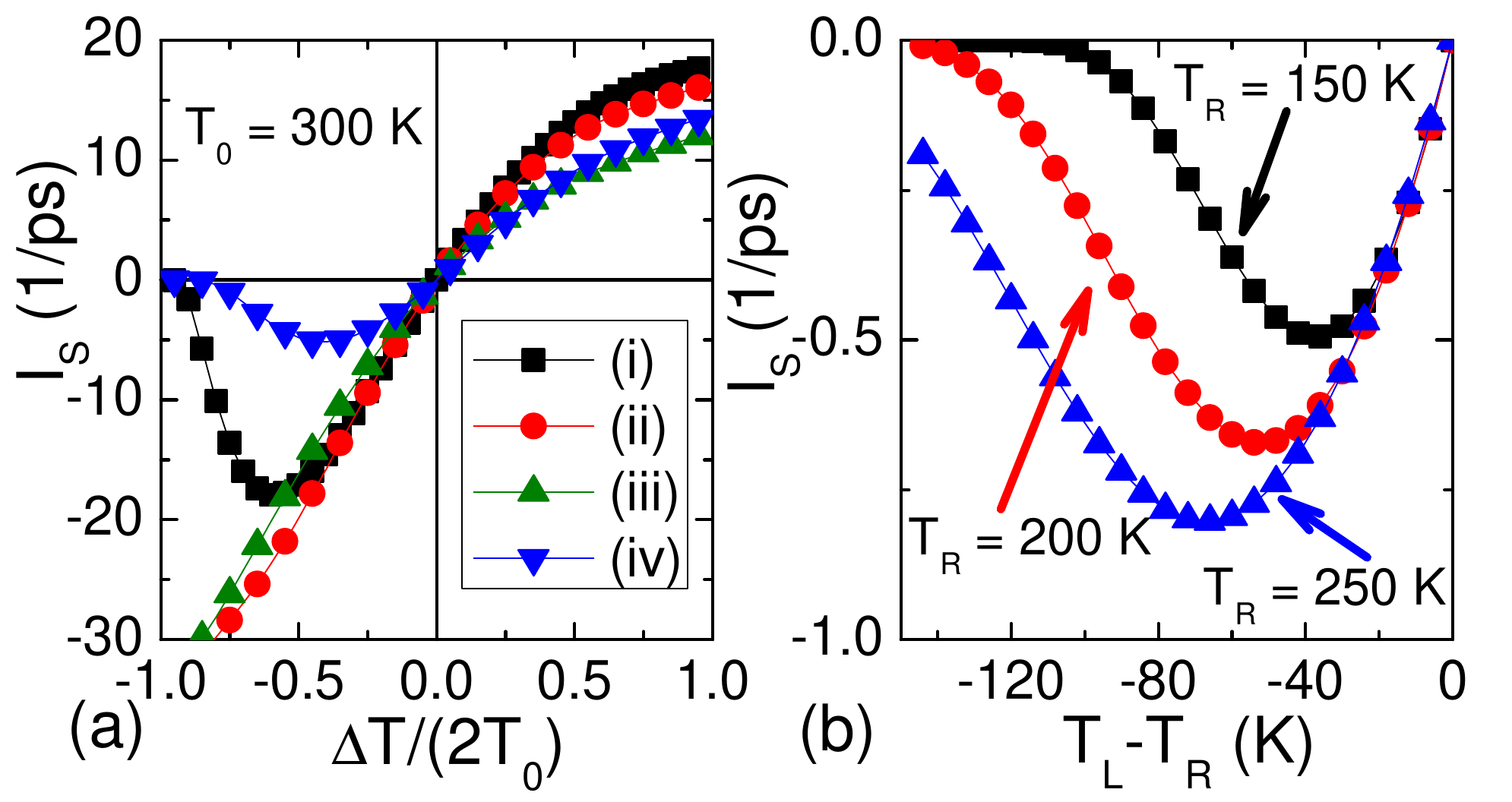}}
\vspace{-8mm}   
\caption{{\bf~Rectification and negative differential spin Seebeck effect in nanoscale magnetic interface}. (a) Magnon spin current $I_S$ as a function of the normalized  temperature bias $\Delta{T}/(2T_0)$ with $T_{L,R}=T_0\pm\Delta{T}/2$, for varying $\varepsilon_{\uparrow,\downarrow}$: (i) $\varepsilon_{\downarrow}=50$~meV, $\varepsilon_{\uparrow}=10$~meV; (ii) $\varepsilon_{\downarrow}=50$~meV, $\varepsilon_{\uparrow}=0$; (iii) $\varepsilon_{\downarrow}=50$~meV, $\varepsilon_{\uparrow}=-25$~meV; (iv) $\varepsilon_{\downarrow}=-25$~meV, $\varepsilon_{\uparrow}=-75$~meV. 
 (b) Magnonic spin Seebeck current as a function of the temperature bias: For $T_R=150$~K, $\varepsilon_{\downarrow}=50$~meV, $\varepsilon_{\uparrow}=30$~meV; For $T_R=200$~K, $\varepsilon_{\downarrow}=60$~meV, $\varepsilon_{\uparrow}=40$~meV; For $T_R=250$~K, $\varepsilon_{\downarrow}=70$~meV, $\varepsilon_{\uparrow}=50$~meV.  We set $\Gamma_J=2.5$~meV and $S_0=16$, which is comparable with the experimental ones in typical magnetic insulators, cf. Refs.~\onlinecite{Adachi2011PRB,para1,para2}.
} \label{fig3}
\end{figure}

For example, considering the typical Lorentzian-type DOS $\rho_L(\varepsilon)=\frac{1}{\pi}\frac{\Gamma}{(\varepsilon-\varepsilon_0)^2+\Gamma^2}$ where the half-width $\Gamma$ is small, the resonant peak of DOS becomes sharp  at $\varepsilon_0$, so that Eq.~(\ref{eq:bulkIS}) reduces to
\begin{eqnarray}
I_S&=&\frac{2S_0}{\hbar}\int^{\infty}_0 d\omega F_R(\omega)\mathcal W(\varepsilon_0, \omega).
\label{eq:bulkIS3}
\end{eqnarray}
It is straightforward to verify that now the rectification and negative differential SSE are present, as illustrated in Fig.~\ref{fig2}. Clearly, the spin current profile is asymmetric under reversing thermal bias -- so called rectification of SSE, and when $|\Delta{T}|\gg40$~K, the further increasing thermal bias gives an anomalously decreasing spin current -- named negative differential SSE. These behaviors are qualitatively similar to those of Eq.~(\ref{eq:IS}) for a nanoscale magnetic interface as plotted in Fig.~\ref{fig3}, which we will discuss in detail later. In fact, for the Pt/YIG junction used in the spin Seebeck measurement~\cite{Uchida2010NatureMat, Uchida2010APL}, the electron DOS of the bulk Pt is strongly fluctuated and is indeed well-described by Lorentzian shapes near the Fermi energy~\cite{Ptbook1, Ptbook2}. 

Even for good metals with smooth DOS, when they are engineered into low-dimensional nanoscale, the electron DOS could become non-smooth and vary strongly in energy due to the quantum confinement effect and other size-induced many-body interactions. Therefore, for low-dimensional nanoscale magnetic interfaces, we can also retain the intriguing properties of rectification and negative differential SSE. For example, the electronic states in one-dimensional (1D) and 2D tight-binding models possess sharp peaked DOS \cite{TBdos}, see also, the strongly-fluctuated DOS of carbon nanotubes~\cite{carbonDOS}. In fact, although the good metal Au has a constant DOS near the Fermi energy for bulk gold~\cite{Ptbook1}, when scaled down to the 1D gold chain, its electronic DOS is peaked as Lorentzian shapes~\cite{1Dgold}. We also note that the longitudinal SSE was recently measured in a thin-film (10nm) Au/YIG (and Pt/YIG) interface junction~\cite{Kikkawa2013PRL}. In principle, we can further reduce the thickness of the metallic thin-film Au (or Pt) so that electrons will be confined in 2D and the DOS will be step-like functions. As such, we also expect to retain the rectification and negative differential SSE in such setup when we are far from the small temperature bias regime. In fact, for the 2D free electron gas, we can obtain its DOS as a step function $\rho_L(\varepsilon)=C_L\Theta(\varepsilon-\mu_0)$. As such, Eq.~(\ref{eq:bulkIS}) reduces to $I_S=\frac{2S_0C_L}{\hbar}\int{d}\omega{F}_{R}(\omega)k_BT_L\ln\frac{2e^{\hbar\omega/(k_BT_L)}}{e^{\hbar\omega/(k_BT_L)}+1}\left[N_{L}(\omega)-N_{R}(\omega)\right]$. It is easy to check that this case retains the rectification and negative differential SSE, similar to the behaviors in Fig.~\ref{fig2}.

Let us finally exemplify the rectification and negative differential SSE in a 0D interface as in the situation of quantum point contacts. As such, the electron states near the interface can be simplified as local states on a two-level quantum dot: $H_C=\varepsilon_{\uparrow}d^{\dag}_{\uparrow}d_{\uparrow}+\varepsilon_{\downarrow}d^{\dag}_{\downarrow}d_{\downarrow}$, where $d^{\dag}_{\sigma}(d_{\sigma})$ is the creation (annihilation) operator of the local electron with spin $\sigma$ and energy $\varepsilon_{\sigma}$ at the interface [see Fig.~\ref{fig1}(b)]. 
The local electron is freely exchanged with the electron reservoir $H_L$ through the coupling $V_L=\sum_{k\sigma}t_{k\sigma}c^{\dag}_{k\sigma}d_{\sigma}+H.c.$. The exchange coupling to the right magnon reservoir $H_R$ is described by~\cite{Takahashi2010JPCS, Adachi2011PRB, Zhang2012PRB,Sothmann2012EPL}
$V_{R}=-\sum_q J_q[S^{z}_q(d^{\dag}_{\uparrow}d_{\uparrow}-d^{\dag}_{\downarrow}d_{\downarrow})+S^{-}_qd^{\dag}_{\uparrow}d_{\downarrow}+S^{+}_qd^{\dag}_{\downarrow}d_{\uparrow}]$.
It is clear that the first term just splits the two local energy levels of the interface so that the contribution can be absorbed into a renormalized $\varepsilon_{\sigma}$ and we have the splitting $\varepsilon_{\downarrow}>\varepsilon_{\uparrow}$ generally. In practice, multiple point contacts can be sandwiched between metal and insulating magnets so that multiple transport channels will form in parallel to enhance the transport signal and efficiency [see Fig.~\ref{fig1}(c)]. Without loss of generality, we only focus on the single channel case, as an in-principle demonstration of the rectification and negative differential SSE. 

Following the same approach as above~\cite{Supple}, we obtain the spin current for the 0D interface system:
\begin{equation}
I_S=\frac{2 S_0}{\hbar}\Gamma_J \big[f_{L\downarrow}(1-f_{L\uparrow})(1+N_R)-f_{L\uparrow}(1-f_{L\downarrow})N_R\big],
\label{eq:IS} 
\end{equation}
where $f_{L\sigma}=[e^{{(\varepsilon_{\sigma}-\mu_{\sigma})}/{(k_BT_L)}}+1]^{-1}$, $N_R=[e^{(\varepsilon_{\downarrow}-\varepsilon_{\uparrow})/{(k_BT_R)}}-1]^{-1}$, and $\Gamma_J=2\pi\sum_q J^2_q \delta(\varepsilon_{\downarrow}-\varepsilon_{\uparrow}-\hbar\omega_q)$ can be regarded as a constant in the wide-band limit. The expression is similar to the spin current of the Lorentzian-DOS case in Eq.~(\ref{eq:bulkIS3}) and they share qualitatively the same behaviors. The mere difference is that instead of an integral over all magnon spectra there, the spin current here only contains a single-mode resonant transfer ($\hbar\omega_q=\varepsilon_{\downarrow}-\varepsilon_{\uparrow}$), which is imposed by the delta function $\delta(\varepsilon_{\downarrow}-\varepsilon_{\uparrow}-\hbar\omega_q)$. 

The rectification of SSE is clearly displayed in Fig.~\ref{fig3}(a) where the thermal-induced spin currents are asymmetric with respect to reversing the temperature bias. In particular, when the two levels are either both above the chemical potential $\varepsilon_{\downarrow}>\varepsilon_{\uparrow}>\mu_0$ [see case (i) in Fig.~\ref{fig3}(a)] or both below $\mu_0>\varepsilon_{\downarrow}>\varepsilon_{\uparrow}$ [see case (iv) in Fig.~\ref{fig3}(a)], we can have the negative differential SSE: When increasing the temperature difference $|T_L-T_R|=|\Delta{T}|$, instead of observing an increasing spin current as in linear response regime, we get a decreasing and even vanishing spin current from the right magnetic insulator to the left metallic lead. It is readily to prove that when $\varepsilon_{\downarrow}\geqslant\mu_0\geqslant\varepsilon_{\uparrow}$, $\partial{I}_S/\partial\Delta{T}$ is always positive so that the negative differential SSE is absent in this parameter regime, as shown by cases (ii) and (iii) in Fig.~\ref{fig3}(a).
Figure~\ref{fig3}(b) shows that the negative differential SSE also exists for a wide range of temperatures. 

The emergence of negative differential SSE can be reasoned as follows: when $T_R>T_L$, the thermal bias drives spin current from the right to the left. If near the linear response regime, it is natural to have a positive differential SSE that lowering $T_L$ will increase the thermal bias which in turn increases the spin current. If we further decrease $T_L$ with assuming two levels  are both above (below) the chemical potential, the two states will be both depleted (occupied), which in turn severely suppresses  the magnon emission/absorption process that requires the concurrence of one occupied and one empty state. As a consequence, the Seebeck spin conductance will decrease although the thermal bias increases. When the conductance decreases faster than the increasing of the bias, the negative differential SSE emerges. 

In summary, we have uncovered the rectification and negative differential SSE in the metal-insulating magnet interface, as a consequence of the strongly-fluctuated electron DOS of the metallic lead. We then have exemplified the identification of these intriguing spin Seebeck properties in several typical cases. Since the magnon carries not only spin but also energy, we additionally possess the rectification and negative differential thermal conductance, as in dielectric phononics~\cite{phononics}. 

Note that throughout the work, we have ensured the condition $2S_0\gg\langle{a}^{\dag}a\rangle$ so that the noninteracting magnon picture assumed in the theory is valid. In fact, in a following work we have shown that the high-order magnon-magnon interaction even becomes crucial for the manifestation of asymmetric and negative differential SSE in magnon tunneling junctions~\cite{RenSSE}, based on  which the concept of a functional spin Seebeck transistor is illustrated. If considering the ferromagnetic-paramagnetic phase transition, the rectification and negative differential SSE will be enhanced. 
Our findings can also be readily generalized to the interfaces of metal-magnetic metal/semiconductor or the ferromagnetic metal-magnetic insulator interfaces.
Since recent studies imply the important role of phonon-drag in SSE \cite{Xiao2010PRB, Adachi2010APL, Jaworski2011PRL, Uchida2011NatureMater, Jaworski2012Nature}, taking account of the effect of nonequilibrium phonons on rectification and negative differential SSE would be an interesting topic. By integrating the phononics~\cite{phononics} with spintronics~\cite{spintronics}, magnonics~\cite{magnonics} and spin caloritronics~\cite{BauerReview}, we expect to invigorate more opportunities to achieve the smart control of energy and information in low-dimensional nanodevices.

\begin{acknowledgments}
{J.R. acknowledges the support from National Nuclear Security Administration of the U.S. DOE at LANL under Contract No. DE-AC52-06NA25396 through the LDRD Program.}
\end{acknowledgments}


\begin{thebibliography}{300}
\bibitem{phonon_function} Generally, the Eliashberg-like function $F_R(\omega)$ also has electronic energy dependence \cite{Mahanbook}. This is because $F_R(\omega)$ contains both the magnon DOS and the electron-magnon exchange couplings, the latter of which also depends on electron energy. By considering strong electronic-energy-dependent $F_R(\varepsilon, \omega)$ in some designed magnetic interfacial systems, we are able to have similar rectification and negative differential transport effects.
\end{thebibliography}


\newpage

\setcounter{figure}{0}
\setcounter{equation}{0}

\renewcommand\thefigure{S\arabic{figure}}
\renewcommand\theequation{S\arabic{equation}}
\renewcommand\bibnumfmt[1]{[S#1]}
\renewcommand{\citenumfont}[1]{S#1}

\begin{widetext}

%
\begin{center}
{\textbf{Supplemental Material for \\``Predicted Rectification and Negative Differential Spin Seebeck Effect at Magnetic Interfaces"}}
\end{center}

\vspace{0.5cm}

In this supplementary material, I provide the detailed derivation of the spin Seebeck transport across metal-magnetic insulator interfaces, which readily lead to the spin and heat currents Eqs.~(\ref{eq:bulkIS}, \ref{eq:bulkIQ}, \ref{eq:W}) in the main text. The thermal-spin transport in the two-level nanoscale interface is also briefly derived, which leads to Eq.~(\ref{eq:IS}) in the main text.

\section{Spin and Heat Currents across Macroscopic magneitc Interface}

Considering each magnon carries an unit spin angular momentum of $-\hbar$ (associated with a magnetic moment), the magnonic spin current is equivalent to a spin-down current, which can be obtained by the Heisenberg equation $I_S=\frac{i}{\hbar}\langle[H_{sd},\sum_qa^{\dag}_qa_q]\rangle$. Thus, by substituting $H_{sd}$ we can get the magnonic spin current as:
\begin{equation}
I_S=\frac{i}{\hbar}\sum_{k,q} J_q \left(\langle  S^-_qc^{\dag}_{k\uparrow}c_{k+q\downarrow} \rangle -\langle  S^+_qc^{\dag}_{k+q\downarrow}c_{k\uparrow} \rangle\right).
\label{eq:Isd}
\end{equation}
Notice that the expectation value $\langle  S^-_qc^{\dag}_{k\uparrow}c_{k+q\downarrow} \rangle$  is just the complex conjugate of $\langle  S^+_qc^{\dag}_{k+q\downarrow}c_{k\uparrow} \rangle$, as a next step, one has only to evaluate the equation of motion:
\begin{eqnarray}
&&\frac{d}{dt}\langle  S^+_qc^{\dag}_{k+q\downarrow}c_{k\uparrow} \rangle =\frac{i}{\hbar} \langle  [H_L+H_{sd}+H_R, S^+_qc^{\dag}_{k+q\downarrow}c_{k\uparrow}] \rangle  \\
&=& \frac{i}{\hbar}(\varepsilon_{k+q\downarrow}-\varepsilon_{k\uparrow}-\hbar\omega_q)\langle  S^+_qc^{\dag}_{k+q\downarrow}c_{k\uparrow} \rangle+
\frac{i}{\hbar}J_q\langle[S^+_qc^{\dag}_{k+q\downarrow}c_{k\uparrow}, S^-_qc^{\dag}_{k\uparrow}c_{k+q\downarrow}]\rangle. \nonumber
\end{eqnarray}
Note that the argument $t$ of each operator is hidden for clarity. Following Ref.~\cite{sNEGFbook}, in the steady state of large $t$ limit, one can write $\langle  S^+_qc^{\dag}_{k+q\downarrow}c_{k\uparrow}S^-_qc^{\dag}_{k\uparrow}c_{k+q\downarrow}\rangle=\langle  S^+_qS^-_q \rangle \langle c^{\dag}_{k+q\downarrow}c_{k+q\downarrow} \rangle \langle c_{k\uparrow}c^{\dag}_{k\uparrow} \rangle=2S_0[1+N_{R}(\omega_q)]f_{L\downarrow}(\varepsilon_{k+p})[1-f_{L\uparrow}(\varepsilon_{k})]$ where $N_R(\omega_q)=[e^{{\hbar\omega_q}/{(k_BT_R)}}-1]^{-1}$ is the magnon distribution function in the right lead, which acts as a thermal bath with temperature $T_R$; $f_{L\sigma}(\varepsilon_k)=[e^{{(\varepsilon_{k}-\mu_{\sigma})}/{(k_BT_L)}}+1]^{-1}$ denotes the electron distribution function at the central dot, which is in equilibrium with the left lead at temperature $T_L$.  

Note since we are considering the metal lead without magnetic order in our present work, the energy spectra of the different spin degree of freedom are the same. In this way, we ignore the spin subscript  for $\varepsilon_k$ and $\varepsilon_{k+q}$. When extend to the ferromagnetic metal case, we need keep the subscript as $\varepsilon_{k\uparrow}$ and $\varepsilon_{k+q\downarrow}$.

Similarly, $\langle  S^-_qc^{\dag}_{k\uparrow}c_{k+q\downarrow} S^+_qc^{\dag}_{k+q\downarrow}c_{k\uparrow} \rangle=\langle  S^-_q S^+_q \rangle \langle c_{k+q\downarrow}c^{\dag}_{k+q\downarrow} \rangle \langle c^{\dag}_{k\uparrow}c_{k\uparrow} \rangle=2S_0 N_R(\omega_q) f_{L\uparrow}(\varepsilon_k)[1-f_{L\downarrow}(\varepsilon_{k+q})]$. Thus, by applying the Markov condition in the steady state $t\rightarrow\infty$, one obtains: 
\begin{equation}
\langle  S^+_qc^{\dag}_{k+q\downarrow}c_{k\uparrow} \rangle=\frac{i}{\hbar}2S_0J_q\int^{\infty}_0d\tau e^{i(\varepsilon_{k+q}-\varepsilon_{k}-\hbar\omega_q)\tau/\hbar}\left\{f_{L\downarrow}(\varepsilon_{k+q})[1-f_{L\uparrow}(\varepsilon_k)][1+N_R(\omega_q)]-f_{L\uparrow}(\varepsilon_k)[1-f_{L\downarrow}(\varepsilon_{k+q})]N_R(\omega_q)\right\}.
\end{equation} 
Substituting it into Eq. (\ref{eq:Isd}), we finally arrive at the expression of spin current:
\begin{equation}
I_S=\frac{2S_0}{\hbar}\int^{\infty}_0 d\omega F_R(\omega) \int^{\infty}_{-\infty} d\varepsilon \rho_L(\varepsilon) \left\{f_{L\downarrow}(\varepsilon+\hbar\omega)[1-f_{L\uparrow}(\varepsilon)][1+N_R(\hbar\omega)]-f_{L\uparrow}(\varepsilon)[1-f_{L\downarrow}(\varepsilon+\hbar\omega)]N_R(\hbar\omega)\right\},      \label{Seq:bulkIS}
\end{equation}
where $f_{L\sigma}(\varepsilon)=[e^{{(\varepsilon-\mu_{\sigma})}/{(k_BT_L)}}+1]^{-1}$ is the electron distribution with spin $\sigma$ in the left metal side that is equilibrium at temperature $T_L$ and $N_R(\hbar\omega)=[e^{{\hbar\omega}/{(k_BT_R)}}-1]^{-1}$ denotes the magnon distribution in the right magnetic insulator that is equilibrium at temperature $T_R$; $\rho_L(\varepsilon)$ denotes the electron DOS in the left metal side; $F_R(\omega)$ contains the magnon DOS and the electron-magnon coupling strength, which is reminiscent of Eliashberg function~\cite{sMahanbook} in the field of electron-phonon scattering.
Generally, the Eliashberg-like function $F_R(\omega)$ also has electronic energy dependence \cite{sMahanbook}, as $F_R(\varepsilon,\omega)=2\pi\sum_q J_q^2\delta(\omega-\omega_q)\delta(\varepsilon-\varepsilon'+\hbar\omega_q)=2\pi\int  d\omega_q \rho_R(\omega_q)J_{\omega_q}^2\delta(\omega-\omega_q)\delta(\varepsilon-\varepsilon'+\hbar\omega_q)=2\pi \rho_R(\omega) J_{\omega}^2\delta(\varepsilon-\varepsilon'+\hbar\omega)$.

Since magnon carries energy, we can also calculate the magnonic heat current through  $I_Q=\frac{i}{\hbar}\langle[H_{sd},\sum_q\hbar\omega_qa^{\dag}_qa_q]\rangle$, which readily leads to

\begin{equation}
I_Q=\frac{2S_0}{\hbar}\int^{\infty}_0 d\omega F_R(\omega) \hbar\omega \int^{\infty}_{-\infty} d\varepsilon \rho_L(\varepsilon) \left\{ f_{L\downarrow}(\varepsilon+\hbar\omega)[1-f_{L\uparrow}(\varepsilon)][1+N_R(\hbar\omega)]-f_{L\uparrow}(\varepsilon)[1-f_{L\downarrow}(\varepsilon+\hbar\omega)]N_R(\hbar\omega) \right\}.   
\label{Seq:bulkIQ}
\end{equation}

Let us examine the Onsager reciprocal relation~\cite{sSothmann2012EPL,sJohnson1987PRB,sKovalev2012EPL,sRen2012PRB} for the thermal-spin transport coefficients in the linear response regime.
Considering $\mu_{\downarrow,\uparrow}=\mu_0\pm\Delta\mu_s/2$, $T_{L, R}=T_0\pm\Delta T/2$,  we expand the expressions of spin and heat currents to the first order of spin voltage and thermal bias ($\Delta\mu_s,\Delta{T}\rightarrow0$), yielding
\begin{eqnarray}
\binom{I_S}{I_Q}=
\left(
\begin{array}{cc}
\mathcal L_0  &  \mathcal L_1  \\
\mathcal L_1  &   \mathcal L_2
\end{array}
\right)
\binom{\Delta \mu_s}{{\Delta T}/{T_0}},
\end{eqnarray}
with 
$\mathcal{L}_n=\frac{S_0}{\hbar}{\int}d\omega F_R(\omega){\int}d\varepsilon\rho_L(\varepsilon)\frac{(\hbar\omega)^n\text{csch}\frac{\hbar\omega}{2k_BT_0}\text{sech}{\frac{\varepsilon-\mu}{2k_BT_0}}\text{sech}{\frac{\varepsilon+\hbar\omega-\mu}{2k_BT_0}}}{4k_BT_0}$. Clearly, the Onsager relation is satisfied.
$\mathcal{L}_0=\partial_{\Delta \mu_s}I_S|_{\Delta{T}=0}$ is the spin conductance under the spin voltage. The spin Seebeck coefficient is $\mathcal{L}_1/(\mathcal{L}_0T_0)=-\Delta \mu_s/\Delta T|_{I_S=0}$, depicting the power of generating spin voltage by the temperature bias.
The spin Peltier coefficient is $\mathcal{L}_1/\mathcal{L}_0=I_Q/I_S|_{\Delta T=0}$, depicting the power of heating or cooling carried by per unit spin current.
In this linear response regime, the spin Seebeck and Peltier coefficients are symmetric when reversing $\Delta{T}\rightarrow-\Delta{T}$ and $\Delta\mu_s\rightarrow-\Delta\mu_s$. However, as we can see in the main text, when we go to the nonlinear response regime ($\Delta\mu_s,\Delta{T}\gg0$), the rectification of SPE will be present and the rectification of SSE will emerge conditionally. In some cases, we can even have the negative differential SSE. 

\section{Spin and Heat Currents in Nanoscale Magnetic Interface}

Considering each magnon carries an angular momentum of $-\hbar$ (associated with a magnetic moment), the magnon current is equivalent to a spin-down current, which is then obtained by the Heisenberg equation of motion $I_S:=\frac{i}{\hbar}\langle[\frac{d^{\dag}_{\downarrow}d_{\downarrow}-d^{\dag}_{\uparrow}d_{\uparrow}}{2}, V_R]\rangle$ or $I_S:=\frac{i}{\hbar}\langle[V_R, \sum_qa^{\dag}_qa_q]\rangle$. From either definition, we can get the same spin (magnon) current:
\begin{equation}
I_S=\frac{i}{\hbar}\sum_q J_q \left(\langle  S^-_qd^{\dag}_{\uparrow}d_{\downarrow} \rangle -\langle  S^+_qd^{\dag}_{\downarrow}d_{\uparrow} \rangle\right).
\label{eq:Isd2}
\end{equation}
Notice that the expectation value $\langle  S^-_qd^{\dag}_{\uparrow}d_{\downarrow} \rangle$  is just the complex conjugate of $\langle  S^+_qd^{\dag}_{\downarrow}d_{\uparrow} \rangle$, as a next step, one has only to evaluate the equation of motion:
\begin{eqnarray}
\frac{d}{dt}\langle  S^+_qd^{\dag}_{\downarrow}d_{\uparrow} \rangle =\frac{i}{\hbar} \langle  [H_C+V_R+H_R, S^+_qd^{\dag}_{\downarrow}d_{\uparrow}] \rangle 
= \frac{i}{\hbar}(\varepsilon_{\downarrow}-\varepsilon_{\uparrow}-\hbar\omega_q)\langle  S^+_qd^{\dag}_{\downarrow}d_{\uparrow} \rangle+
\frac{i}{\hbar}J_q\langle[S^+_qd^{\dag}_{\downarrow}d_{\uparrow}, S^-_qd^{\dag}_{\uparrow}d_{\downarrow}]\rangle.
\end{eqnarray}
Note that the argument $t$ of each operator is hidden for clarity. Following the same approach, in the steady state of large $t$ limit, one can write $\langle  S^+_qd^{\dag}_{\downarrow}d_{\uparrow}S^-_qd^{\dag}_{\uparrow}d_{\downarrow} \rangle=\langle  S^+_qS^-_q \rangle \langle d^{\dag}_{\downarrow}d_{\downarrow} \rangle \langle d_{\uparrow}d^{\dag}_{\uparrow} \rangle=2S_0(1+N_R)f_{L\downarrow}(1-f_{L\uparrow})$ where $N_R=[e^{{\hbar\omega_q}/{(k_BT_R)}}-1]^{-1}$ is the magnon distribution function in the right lead, which acts as a thermal bath with temperature $T_R$; $f_{L\sigma}=[e^{{(\varepsilon_{\sigma}-\mu_{\sigma})}/{(k_BT_L)}}+1]^{-1}$ denotes the electron distribution function at the central dot, which is in equilibrium with the left lead at temperature $T_L$. Similarly, $\langle  S^-_qd^{\dag}_{\uparrow}d_{\downarrow} S^+_qd^{\dag}_{\downarrow}d_{\uparrow} \rangle=\langle  S^-_q S^+_q \rangle \langle d_{\downarrow}d^{\dag}_{\downarrow} \rangle \langle d^{\dag}_{\uparrow}d_{\uparrow} \rangle=2S_0 N_R f_{L\uparrow}(1-f_{L\downarrow})$. Thus, by applying the Markov condition in the steady state $t\rightarrow\infty$, one obtains: $\langle  S^+_qd^{\dag}_{\downarrow}d_{\uparrow} \rangle=\frac{i}{\hbar}2S_0J_q\int^{\infty}_0d\tau e^{i(\varepsilon_{\downarrow}-\varepsilon_{\uparrow}-\hbar\omega_q)\tau/\hbar}[f_{L\downarrow}(1-f_{L\uparrow})(1+N_R)-f_{L\uparrow}(1-f_{L\downarrow})N_R]$. Substituting it into Eq. (\ref{eq:Isd2}), we finally arrive at the expression of spin current:
\begin{equation}
I_S=\frac{2 S_0}{\hbar}\Gamma_J \big[f_{L\downarrow}(1-f_{L\uparrow})(1+N_R)-f_{L\uparrow}(1-f_{L\downarrow})N_R\big],
\label{eq:IS2}
\end{equation}
where $\Gamma_J=2\pi\sum_q J^2_q \delta(\varepsilon_{\downarrow}-\varepsilon_{\uparrow}-\hbar\omega_q)$. 

This expression is familiar from the results of applying Fermi's golden rule to the definition Eq. (\ref{eq:Isd2}). The first product $f_{L\downarrow}(1-f_{L\uparrow})(1+N_R)$ describes the relaxation rate of the occupied higher spin-down state flipping to the empty lower spin-up state with emitting a magnon with energy $\hbar\omega_q$ into the right lead. The second product $f_{L\uparrow}(1-f_{L\downarrow})N_R$ reversely describes the excitation rate of the occupied lower spin-up state flipping to the empty higher spin-down state with absorbing a magnon from the right lead. The two spin-flip processes, accompanied by
the magnon emission/absorption, conserve not only the angular momentum, but also the energy, which is imposed by the delta function $\delta(\varepsilon_{\downarrow}-\varepsilon_{\uparrow}-\hbar\omega_q)$. In this way, only magnons with energy $\hbar\omega_q=\varepsilon_{\downarrow}-\varepsilon_{\uparrow}$ is able to transfer through the magnetic interface. Therefore, for the energy current carried by magnons, we can also obtain similarly:
\begin{equation}
I_Q\!\!=\!\!\frac{2 S_0}{\hbar}\Gamma_{\!\!J} (\varepsilon_{\downarrow}-\varepsilon_{\uparrow})\! \big[f_{L\downarrow}(1-f_{L\uparrow})(1+N_{\!R})-f_{L\uparrow}(1-f_{L\downarrow})N_{\!R}\big].
\label{eq:IQ2}
\end{equation}

Let us examine the Onsager reciprocal relation~\cite{sSothmann2012EPL,sJohnson1987PRB,sKovalev2012EPL,sRen2012PRB}. Considering $\mu_{\downarrow,\uparrow}=\mu_0\pm\Delta \mu_s/2$, $T_{L, R}=T_0\pm\Delta T/2$, the thermal-spin transport coefficients are conventionally considered in the linear response regime: expanding to the first order of spin voltage bias $\Delta \mu_s=\mu_{\downarrow}-\mu_{\uparrow}$ and thermal bias $\Delta T=T_L-T_R$, which yields
\begin{eqnarray}
\binom{I_S}{I_Q}=
\left(
\begin{array}{cc}
\mathcal G  &  \mathcal G \mathcal S T_0  \\
\mathcal G  \Pi  &   \kappa T_0
\end{array}
\right)
\binom{\Delta \mu_s}{\Delta T/T_0},
\end{eqnarray}
where 
 $\mathcal G=\frac{S_0\Gamma_J}{\hbar T_0}/({\sinh}[\frac{\varepsilon_{\downarrow}-\mu_0}{k_BT_0}]-{\sinh}[\frac{\varepsilon_{\uparrow}-\mu_0}{k_BT_0}]+{\sinh}[\frac{\varepsilon_{\downarrow}-\varepsilon_\uparrow}{k_BT_0}])$
is the spin conductance, generated by the spin voltage difference. $\kappa=\mathcal G(\varepsilon_{\downarrow}-\varepsilon_{\uparrow})^2/T_0$ denotes the heat conductance, produced by the temperature bias. $\mathcal S=-\Delta \mu_s/\Delta T|_{I_S=0}=(\varepsilon_{\downarrow}-\varepsilon_{\uparrow})/T_0$ is the spin Seebeck coefficient, depicting the power of generating spin voltage by the temperature bias. $\Pi=I_Q/I_S|_{\Delta T=0}=\varepsilon_{\downarrow}-\varepsilon_{\uparrow}$ is the spin Peltier coefficient, depicting the power of heating or cooling carried by per unit spin current. It is seen clearly that we have $\Pi=\mathcal S T_0$ so that the Onsager reciprocal relation $\mathcal G \mathcal S T_0=\mathcal G \Pi$ is fulfilled in the present system.

It is worth noting that when the right magnetic insulator is replaced by magnetic semiconductor or metal, the Onsager reciprocal relation will be seemingly violated.  The reason is that itinerant conduction electrons inside the right lead will provide additional hidden pathways for thermal-spin-charge transports, which could just circulate within the right lead and will not contribute to the net transport. If we are unaware of those hidden currents and still consider the thermodynamic conjugate transports simply from the left to the right, we will observe the seemingly violated Onsager relation. Similar situation occurs when the left metallic lead has additional magnetic orders, like ferromagnetic metal. Therefore, to get the correct symmetric reciprocal relation, one need carefully dissect all possible transport pathways, excluding the hidden currents which do not contribute to the net transport and other irrelevant currents. Only in this way, the correct Onsager relations will be recovered. 

In this linear response regime, the spin Seebeck and Peltier coefficients are symmetric when reversing $\Delta \mu_s\rightarrow-\Delta \mu_s$ and $\Delta T\rightarrow -\Delta T$. However, as we can see in the main text, when we go to the nonlinear response regime, the rectification of SPE and SSE will emerge. In some cases, we can even have the negative differential SSE. 

\end{widetext}

\end{document}